\documentstyle[aps,preprint]{revtex}
\begin{document}
\draft
\title{
Nuclear Shell Model by the Quantum Monte Carlo Diagonalization method
}
\author{Michio Honma$^1$, Takahiro Mizusaki$^2$
 and Takaharu Otsuka$^{2,3}$}
\address{$^1$Center for Mathematical Sciences, University of Aizu\\
 Tsuruga, Ikki-machi Aizu-Wakamatsu, Fukushima 965, Japan }
\address {$^2$Department of Physics, University of Tokyo, Hongo, Tokyo 113,
Japan}
\address{$^3$RIKEN, Hirosawa, Wako-shi, Saitama 351-01, Japan}
\date {Submitted 27 April 1996}

\maketitle

\begin{abstract}
The feasibility of shell-model calculations is radically extended by 
the Quantum Monte Carlo Diagonalization method 
with various essential improvements.
The major improvements are made  in the sampling for the generation of 
shell-model basis vectors,
 and in the restoration of symmetries such as angular momentum and isospin.
Consequently the level structure of low-lying states can be studied
% by large-scale shell-model calculations
 with realistic interactions.
After testing this method on $^{24}$Mg, we present
first results for energy levels and
 $E2$ properties of $^{64}$Ge,
 indicating its large and $\gamma$-soft deformation.
\end{abstract}

\pacs{PACS numbers: 21.60.Ka, 21.60.Cs, 24.10.Cn}

\narrowtext

The nuclear shell model has been successful in the description of 
various aspects of nuclear structure, partly 
% because of its least model-dependent nature.
because it is based on a minimum number of natural assumptions.
Although the direct diagonalization of the
 hamiltonian matrix in the full valence-nucleon Hilbert space
 is desired, the dimension of such a space is too large in many cases,
preventing us from performing the full calculations.
The direct diagonalization has been carried out up to 
 $^{48}$Cr \cite{poves}.
Recently, in order to relax this restriction drastically,
 stochastic approaches, for instance, the shell model Monte 
 Carlo (SMMC) method \cite{mcsm}, have been investigated.
In fact, ground-state \cite{langanke}
 and thermal properties \cite{dean} have been
 described well by the SMMC method.
%However, the SMMC method is useful mainly for the ground state properties, 
% and the form of the interaction is limited
% due to the minus-sign problem.

We have presented the Quantum Monte Carlo Diagonalization (QMCD) method
 \cite{qmcd} by utilizing the auxiliary field
 Monte Carlo technique as in the SMMC method,
 but in a quite different way. 
In general, low-lying states of nuclei are 
 described to a good extent in terms of static and/or dynamic mean fields
and their fluctuations.
The basic idea of the QMCD method is  to diagonalize the
 shell model hamiltonian,  by using this property, in a subspace spanned by
a small number of selected basis states
 obtained by stochastically generated one-body fields.
% describing a fluctuation around the mean field.
Thus, the ground state and  several excited states can be obtained.
The QMCD method has been applied
 to the Interacting Boson Model \cite{qmcd,mproj}.
In this Letter the QMCD method is
 revised considerably in various aspects so as to be capable of
 performing large-scale shell model calculations
 with realistic nuclear forces.
As examples, $^{24}$Mg and $^{64}$Ge are taken.
In particular, $^{64}$Ge is an $N=Z$ proton-rich unstable nucleus
 manifesting a $\gamma$-soft structure,
 with a wide range of theoretical interpretations
 ( see ref.\cite{ge} ).
Thus, the shell model calculation can play a crucial role
 for clarifying the level structure,
 but so far such attempts have been impossible
 due to the large dimension ($\sim 1\times 10^9$).

We first sketch the QMCD process very briefly,
  referring to relevant equations of ref.\cite{qmcd}.
More details on certain basic points can be found in ref.\cite{qmcd}.
The shell model hamiltonian consisting of
 single particle energies and a two-body interaction
 can be written in the quadratic form of $N_f$
 one-body operators $O_{\alpha}$:
\begin{equation}
H = \sum_{\alpha=1}^{N_{f}} (E_{\alpha} O_{\alpha} + \frac{1}{2} 
V_{\alpha} O_{\alpha}^{2}).
\label{eqn:decomp}
\end{equation}
We consider the imaginary time evolution operator 
 with $N_t$ slices: 
$e^{-\beta H}=\prod\nolimits_{n=1}^{N_t} {e^{-\Delta \beta H}}$,
 where $\Delta\beta = \beta/N_{t}$.
By applying the Hubbard-Stratonovich transformation 
 at each time slice \cite{hub,suzuki},
 $e^{-\beta H}$ can be expressed as the integral of an operator,
$\prod_{n=1}^{N_{t}} e^{-\Delta \beta h(\vec{\sigma}_{n})}$,
 over $N_f \times N_t$ auxiliary fields $\sigma_{\alpha n}$
 (see Eq.(4) of ref.\cite{qmcd})
 with the Gaussian weight factor
$G(\sigma) = \exp\left ( 
{-\sum_{\alpha, n} \Delta \beta/2 \mid\!V_{\alpha} 
\mid\!\sigma_{\alpha n}^{2}} \right )$.
The one-body hamiltonian $h(\vec{\sigma}_{n})$ is defined by
\begin{equation}
h(\vec{\sigma}_{n}) = \sum_{\alpha} (E_{\alpha} + s_{\alpha} V_{\alpha} 
\sigma_{\alpha n} ) O_{\alpha},
\label{eqn:oneham}
\end{equation}
 where $s_{\alpha}=\pm 1$ \ ($=\pm i$) if $V_{\alpha}<0$ \ ($>0$).
In the QMCD method, by generating a new set of values for
$\sigma=\{\sigma_{\alpha n}\}$ stochastically
 according to $G(\sigma)$,
 a new many-body state is created as,
\begin{equation}
\mid\!\Phi (\sigma) \rangle \propto \prod_{n=1}^{N_{t}} e^{-\Delta \beta 
h(\vec{\sigma}_{n})} \mid\!\Psi^{(0)} \rangle,
\label{eqn:state}
\end{equation}
 where $\mid\!\Psi^{(0)}\rangle$ is an appropriate initial state.
The Hamiltonian is diagonalized in the Hilbert subspace
 spanned by this state and the basis states previously obtained.
If this new state improves the result of the diagonalization
 sufficiently well, this state is added to the basis states.
The number of such basis sates is
 referred to as the QMCD basis dimension, and is increased
 until reasonable convergence is achieved.

It is convenient to adopt basis states 
 in the form of Slater determinants:
%\begin{equation}
$
%\mid\!\Phi\rangle = \prod_{\alpha=1}^{N}a_\alpha^\dagger \mid\! - \rangle,
\prod_{\alpha=1}^{N}a_\alpha^\dagger \mid\! - \rangle,
$
%\end{equation}
 where $N$ denotes the number of valence nucleons,
 $\mid\! - \rangle$ is an inert spherical core,
 and $a_\alpha^\dagger$ represents the nucleon creation operator
 in a canonical single-particle state $\alpha$, which is a linear combination
of the spherical bases.
Note that,
 if $\mid\!\Psi^{(0)}\rangle$ is a Slater determinant,
 $\mid\!\Phi (\sigma)\rangle$ in Eq.(\ref{eqn:state}) remains
 in the form of a Slater determinant.
%\begin{equation}
% a_\alpha^\dagger = \sum_{i=1}^{N_{sp}} c_i^\dagger D_{i \alpha}.
%\end{equation}
%We can specify the basis state $\mid\!\Phi\rangle$ in terms of 
% an $N_{sp}\times N$ complex matrix $D$.
%Note that the operation of an exponential of any one-body operator
% $T=\sum_{ij}T_{ij}c_i^\dagger c_j$ on a state gives
% a new matrix $D'_{i \alpha} = \sum_{j}T_{ij}D_{j \alpha}$, while
% the form of a Slater determinant remains.

While the QMCD method outlined so far 
 is applicable to fermion systems,
 its capability is limited to simple cases,
 for instance, a single-j model.
Difficulties arise, for example, 
 due to finite single particle energies.
Thus, a substantial further 
 improvement of the method
 is required for realistic shell model calculations.
Such improvements are: (i) the sampling scheme is modified,
 and (ii) additional processes are included  to restore
symmetries.

%This process does not necessarily require many time slices.
%Thus, in the following calculations, we take $N_t=1$
% so as to make the calculation faster.
%The effects of configurations which cannot be reached by $N_t=1$
% can be easily taken into account by increasing the time slices
% or by changing the initial state.
%Figure \ref{fig1} includes the results obtained with $N_t=1$
% by dashed lines,
% otherwise the same parameters are taken as the above results with $N_t=40$.
%Thus, $\beta=20$. 
%One finds a rather remarkable feature that 
%The convergence of the energy and the $J\cdot J$ expectation value is
% somewhat better than the $N_t=40$ calculations.

We start with the sampling. 
In the original version of the QMCD method,
 a rather naive sampling is performed
 (see Eq.(4) of ref.\cite{qmcd}).
This sampling
 creates many unnecessary basis vectors in general, and indeed
 the actual sampling has to be modified for large-scale realistic shell model 
 calculations so that 
 important basis vectors are generated still stochastically but more
 efficiently by considering the many-body dynamics.
%important basis vectors are 
%generated stochastically but based on the many-body dynamics.

The modification regarding the sampling 
 consists of two parts.
In the first part, the basis state generation is refined so as to make use of
the local Hartree-Fock (HF) energy minima.
In the QMCD calculation, one has to generate good basis states
(i.e., Slater determinants in deformed bases)
 which have (i) low values of diagonal matrix elements
 and/or (ii) large off-diagonal matrix elements of the hamiltonian.
%Since we take basis states in the form of Slater determinant,
 The point (i) can be fulfilled by using,
 as $\mid\!\Psi^{(0)}\rangle$ in Eq.(\ref{eqn:state}),
 a deformed HF solution within the present shell model space.
The QMCD process is comprised practically of several segments starting with
 different initial states,
 which are HF states at different local minima.
%The energy of each basis state around a minimum
% then becomes not lower than that of the initial state.
States around a minimum satisfy point (ii) in most cases.
We then rearrange the one-body evolution process
 so that the basis states are sampled most
 frequently near the HF local minima,
accelerating the generation of state vectors having larger overlap with 
eigenstates of interest.
% avoiding generating 
%unnecessary bases.

The hamiltonian is rewritten, by introducing
the constants $c_\alpha$, as
\begin{equation}
H = \sum_\alpha (E_\alpha O_\alpha + \frac{1}{2}
 V_\alpha (O_\alpha - c_\alpha)^2 +
 V_\alpha c_\alpha O_\alpha),
\end{equation}
 where a constant term is omitted.
After the HS transformation, the one-body hamiltonian becomes
\begin{equation}
h(\vec{\sigma}_n) = \sum_\alpha ( (E_\alpha +
 V_\alpha c_\alpha) O_\alpha +
 s_\alpha V_\alpha \sigma_{\alpha n} O_\alpha),
\end{equation}
where the c-number
 $-\sum_\alpha s_\alpha V_\alpha \sigma_{\alpha n} c_\alpha$
 is omitted since it does not change the wave functions apart from
the normalization.
In this expression, the modified one-body term
 $\sum_\alpha (E_\alpha + V_\alpha c_\alpha) O_\alpha$
 includes effects coming from the two-body interaction.
The $c_\alpha$'s  are taken in such a way that this term becomes the HF 
single-particle hamiltonian, $h_{HF}$.
%By taking the values of $c_\alpha$'s 
% to be consistent with a HF state,
%this term becomes its HF single-particle hamiltonian, $h_{\rm HF}$.
With this $h_{\rm HF}$, the QMCD basis state takes the form,
\begin{equation}
\mid\!\Phi (\sigma) \rangle \propto \prod_{n=1}^{N_{t}} e^{-\Delta \beta 
( h_{\rm HF} +  \sum_\alpha s_\alpha V_\alpha \sigma_{\alpha n} O_\alpha)}
 \mid\!\Psi^{(0)} \rangle.
\label{eqn:state1}
\end{equation}
Thus, we simply replace the single particle energy
 $\sum_\alpha E_\alpha O_\alpha$ by $h_{\rm HF}$.
If $|\Psi^{(0)} \rangle$ is the HF state being considered,
 the sampling around $\sigma=0$ generates
 various states around this HF state, including
 Tamm-Dancoff-type states to first order in $\sigma$, and so on.
This treatment is possible for all HF local minima.
%Note that, for $h(\vec{\sigma}_n)$ in Eq.(\ref{eqn:oneham}),
% $\sigma$=0 means no contribution of two-body interaction,
% whereas the sampling is most frequent at $\sigma$=0.

In cases of non-spherical nuclei, many Hartree-Fock local minima 
 appear in the search for the initial state.
By stochastically taking those local minima
 as the initial states,
 it is possible to take into account a wider variety 
 of configurations.

The second major improvement on the sampling
 is the ordering of one-body fields
 according to their importance.
The QMCD method is a method for generating
 favorable basis states for diagonalization,
 and there is no need to carry out
 the stochastic integration over all auxiliary fields.
In constructing the basis states,
 we start with the most relevant part of the hamiltonian,
 which yields fewer auxiliary fields than the
 whole hamiltonian.
The calculation can then be performed more efficiently.
After certain basis states are obtained, we
 take an enlarged portion of the hamiltonian, so that other terms
 of the hamiltonian can be properly included in constructing
 the basis states.
Eventually
 the completeness of the QMCD basis is guaranteed
 for the ground state by taking all fields.

In most cases, the auxiliary fields with large
 values of $\mid\! V_\alpha \mid$ in Eq.(\ref{eqn:decomp})
 turn out to have quadrupole,  hexadecapole or monopole nature.
Therefore it is reasonable to arrange all
 fields in descending order of  $\mid\! V_\alpha \mid$,
 and take them starting from the largest one.
In addition, for a fixed initial state $\mid\!\Psi^{(0)}\rangle$,
 the total strength of each $O_\alpha$ changes 
 due to the Pauli principle and to collective effects.
Therefore we consider an excitation sum-rule;
\begin{equation}
S_\alpha =
\langle \Psi^{(0)}\mid\! O_\alpha^\dagger O_\alpha\mid\! \Psi^{(0)}\rangle
- \mid\! \langle \Psi^{(0)}\mid\! O_\alpha \mid\! \Psi^{(0)}\rangle \mid^2 ,
\end{equation}
 and use it as a practical measure of importance of the  $O_\alpha$'s.
The selection of $O_\alpha$'s according to 
 $\mid\! V_\alpha\mid$ and $S_\alpha$
 plays an essential role in the actual calculations.
%Another elimination is also possible by looking at the
% energy surface with respect to $\sigma$:
%\begin{equation}
%E(\sigma) = \frac{\langle\Phi(\sigma)\mid\! H \mid\!\Phi(\sigma)\rangle}
%{\langle\Phi(\sigma)\mid\!\Phi(\sigma)\rangle}.
%\end{equation}
%This energy surface is rather flat for redundant operators.

Incorporating all the above improvements, the sampling
 is made much more efficient.
Note that this way of sampling clearly differs from that of the SMMC.

We now come to the restoration of symmetries.
We implement explicitly kinematic symmetries
 such as angular momentum and isospin
 into the QMCD method,
 since the restoration of such symmetries proceeds only very slowly 
for wave functions generated stochastically.
In the previous paper \cite{mproj} we have presented the 
 $M$-projection 
 method to restore the magnetic quantum number.

Since a nucleus has rotational symmetry, the restoration of the
 total angular momentum, denoted as $J$, is quite crucial.
In the QMCD method,
 we diagonalize the hamiltonian in the
 laboratory frame by using QMCD bases.
If the QMCD bases contain all components  (i.e., Slater determinants)
 required for the coupling to a good angular momentum,
 the diagonalization restores the rotational symmetry.
% the appropriate superposition of rotated states are automatically
% performed if these states are fully contained in the 
% adopted QMCD basis states.
We accelerate this restoration process,
 by considering rotated states 
 $\exp(-i\theta_y J_y)\exp(-i\theta_z J_z) \mid\!\Phi(\sigma)\rangle$
 as candidates of new basis states.
%Certainly, $M$-projection is carried out properly also \cite{mproj}.
We have found that the restoration of the angular
 momentum is remarkably improved by taking only several
 values of the angle $\theta$'s.
We refer to this method as $J$-drive.
In addition to this,
 the $M$-projection \cite{mproj} is carried out
 for all bases thus created.

We next discuss isospin.
The isospin projection is possible in the same way as the $J$-projection.
In this Letter, however, since we consider only $N$=$Z$ nuclei,
 we keep good isospin in an alternative way.
In the decomposition process,
 Eq.(\ref{eqn:decomp}),
 all one-body operators can be chosen so as to carry a definite
 isospin $T=0$ or $1$ for the isoscalar hamiltonian.
Since the isoscalar fields are dominant over the isovector ones,
 particularly, for $T$=0 states, we start  the QMCD basis generation process
with the isoscalar fields.
%It is found that we can take only isoscalar fields
% as the significant fields for the early stage of the 
% QMCD basis generation process.
Thus, since the initial HF state has $T=0$,
 the isospin is conserved at least until the isovector fields are activated.
It appears that, in $N=Z$ nuclei, 
one obtains sufficiently good results by keeping only the isoscaler fields.
%only the isoscalar fields
% appear to give sufficiently good results.
For $T\neq 0$ states, the isospin projection is definitely needed,
 and results obtained with this procedure  will be presented elsewhere.

As an example of realistic shell model calculations,
 we first consider $^{24}$Mg with the USD interaction \cite{usd}.
Figure \ref{fig1} shows energies and expectation values of 
 $J\cdot J$, for six low-lying states as a function of the
 QMCD basis dimension compared with
 the exact values.
In this case we start with five significant
 fields, and eventually all 144 $T=0$
 one-body operators are activated.
In the process of $J$-drive, three values for $\theta_y$ are employed.
%, and 3 points are tried
% in the process of the $J$-drive.
For 800 QMCD bases, the ground state energy
 becomes $-$86.91 MeV, while the exact value is $-$87.08 MeV.
The dimension of the m-scheme shell-model basis for
 the ground state is 28\,503.
Thus the number of bases is reduced by a factor 1/35
 with a loss of accuracy of only 0.17 MeV in the total energy.
The error due to the truncation of the Hilbert space
 (systematic error) does not exceed two hundred keV
 in the ground-state energy in the present calculations.

Table \ref{table1} shows
 the lowest three energy levels,
 where the QMCD results for 100, 400 and 800 basis dimensions
 are listed together with the exact results.
One finds a remarkable agreement between the QMCD and exact values.
Note that the accuracy of these excitation energies is
 better than that of absolute energies.
In fact the deviations are less than 0.15MeV
 with only 400 basis states. 
In the same table, several $E2$ transition matrix elements
 and quadrupole moments are compared with the exact values.
It can be found that several in-band transition $B$($E2$) values
 are reproduced well with only 100 basis states, and
 other matrix elements are also obtained with 400 basis states.
Thus the QMCD method turns out to be useful,
 especially for the
 study of low-lying collective states.

We now proceed on to full pf shell calculations.
We have confirmed the feasibility of the QMCD method
 by comparing its results with the exact ones \cite{poves}
 for $^{48}$Cr with the KB3 interaction
 \cite{kb3},
 as will be presented elsewhere.
In this Letter, we discuss $^{64}$Ge.
The $m$-scheme dimension of the $M$=0 space is 1\,087\,455\,228,
 which is the second largest one for the $N=Z$ even-even
 pf shell nuclei.
It is larger than the dimension for  $^{48}$Cr 
 by a factor of about 550,
 and the exact diagonalization is hopeless
in the near future.
This nucleus is one of the proton-rich $N=Z$ unstable nuclei, and
 experimental data \cite{ge} suggest that it is $\gamma$-soft.
Thus, it is quite interesting to investigate whether
 we can reproduce such a structure
 by using a realistic interaction, the validity of which has been examined
 at least for the lower part of the pf shell.
In this Letter, we adopt the FPD6 interaction \cite{brown}.
This interaction is derived by fitting experimental data 
in the mass range 41-49,
 and is suggested to be suitable for
 describing nuclei in the upper pf shell \cite{kb3dame}.

%We have no well-established effective interaction for the 
% upper part of the pf shell.
%Thus it is to be carefully examined whether this interaction 
% is applicable to heavier nuclei.
%We have also tried Kuo-Brown \cite{kb} interaction and its 
% modified one (KB3) \cite{kb3}, and found that these interactions
% do not give better description of low-lying 
% excitation spectrum than the present one, especially for $^{56}$Ni.
%This point will be discussed elsewhere.

In Fig.\ref{fig2}, calculated low-lying spectra
 are compared with experimental data.
It is remarkable that the calculated levels
 show a rather good agreement with experiment
 without any adjustment.
The $\gamma$-soft nature is also evident in the calculation.
The calculated ratio of excitation energies of $2^+_2$ to 
 $2^+_1$ is 1.9 and that of $4^+_1$ to $2^+_1$ is 2.6.
Experimentally these ratios are 1.75 and 2.27, respectively. 
The relative magnitudes of $B$($E2$) values are shown in Fig.2.
With $e_p=1.33e$ and $e_n=0.64e$, 
 $B$($E2$;$2^+_1\rightarrow 0^+_1$)=$5\times 10^2$
($e^2$fm$^4$) is obtained, which corresponds to $\beta_2\sim 0.28$.
The $B$($E2$) values of the $4^+_1\rightarrow 2^+_1$ 
 and $2^+_2\rightarrow 2^+_1$
 transitions are about 1.3 times larger than that of 
 $B$($E2$;$2^+_1\rightarrow 0^+_1$), suggesting $\gamma$-softness.
%The $E2$ transition matrix elements
% are shown in Table \ref{table2}.
We obtain 
 $B$($E2$;$2^+_2\rightarrow 0^+_1$)/$B$($E2$;$2^+_2\rightarrow 2^+_1$)
 $\sim$ 2$\times 10^{-3}$, which is quite small
 similarly to the experimental value,
 suggesting $\gamma \sim 30^{\circ}$ in triaxial 
deformation models \cite{ge}.
Calculated quadrupole moments appear to be small
 (typically $\mid\! Q\mid$ $<$ 10$e$fm$^2$), 
 consistently with $\gamma$-softness.
%The deformation parameter calculated from
% $B$($E2$;$0^+_1\rightarrow 2^+_1$)
% is $\beta\sim 0.27$.
%Thus, the present result suggests that $^{64}$Ge is a particularly 
%interesting nucleus with large and triaxial deformation.

The convergence of the results in Fig.2 has been examined by several
calculations with different stochastic parameters.
The typical deviation among different calculations (statistical error)
is about 100 keV for the $2^+_2$ energy level, for instance.
The discrepancy between theoretical and experimental results comes
 partly from the systematic and statistical errors
 in the present method, and partly from the interaction.
The former one is being reduced by improving the method.

Typical occupation numbers of
 $f_{7/2}$, $p_{3/2}$, $p_{1/2}$ and $f_{5/2}$ orbits are
 15.1, 2.6, 0.8 and 5.5, respectively, for low-lying states.
We can see that more than six nucleons 
 are excited from the ($f_{7/2}$)$^{16}$ ($p_{3/2}$)$^8$
 configuration, and that even $f_{7/2}$ is active.
One sees that all these four orbits are mixed.
Because of the huge basis dimension mentioned before,
 the conventional shell model diagonalization is impossible.

The QMCD method can generate, in principle, all basis states
 which are needed to describe the exact eigenstate.
It is free of the assumption of some specific
 collective coordinates as in the usual Generator Coordinate Method.
In addition we can take into account various states
 around many different local energy minima,
 which is difficult in variational approaches 
 with multi-Slater determinants or multi HFB states.

In summary, it has been shown that 
 large-scale realistic shell model calculations
 can be carried out by the QMCD method.
The QMCD method has been improved considerably with respect to
 (1) the sampling of auxiliary fields based on the local energy minima,
 (2) the selection of dominant fields,
 and (3) the explicit implementation of kinematic symmetry requirements.
Several low-lying states of large systems have been described
 in terms of small numbers of QMCD basis states
 with the accuracy of several hundred keV in total energies.
The accuracy of excitation energies and $E2$ transition
 matrix elements is much better.
Such capability of describing low-lying states
 is the major advantage of the present method over the SMMC method.
The present results demonstrate that 
 the shell model calculations with full valence shell configurations
 have become feasible by the QMCD method,
 shading light upon the structure of nuclei
 even beyond the pf shell with more direct relation to
 the effective nucleon-nucleon interaction.
The minus-sign problem seems to be absent in the
 QMCD method, and, hence,
 any effective two-body interaction can be used as it is.

We acknowledge Professors B. A. Brown and A. Poves 
 for providing relevant two-body matrix elements.
We are grateful to Drs. W. Bentz and A. Gelberg for reading the manuscript.
This work was supported in part by Grant-in-Aid for Scientific Research on
Priority Areas (No. 05243102) from the Ministry of Education, Science
and Culture.
A part of this work was carried out on the VPP500 computer at RIKEN as
a part of the Computational Nuclear Physics Project of RIKEN.

\begin{table}
\caption{Comparison between the QMCD and the exact results 
for excitation energies (MeV),
 $B$($E2$) ($e^2$fm$^4$) and quadrupole moments ($e$fm$^2$).
 The effective charges $e_p + e_n = 1.78e$ are used
 \protect \cite{usd}.
 \label{table1}}
\begin{center}
\begin{tabular}{ccrrrcr}
 \multicolumn{2}{c}{observable} & \multicolumn{3}{c}{QMCD dimension} & &
 exact \ \  \\ \cline{3-5}
  & & \ \ \ 100\ \ \ 
 & \ \ \ 400\ \ \  & \ \ \ 800\ \ \  & & \\ \hline
$E_x(2^+_1)$ & &  1.50 & 1.54 & 1.53 & & 1.51 \\
$E_x(2^+_2)$ & &  4.33 & 4.23 & 4.18 & & 4.12 \\
$E_x(4^+_1)$ & &  4.54 & 4.50 & 4.46 & & 4.37 \\ \hline
$B$($E2$;$2^+_1\rightarrow 0^+_1$) & & 74.1 & 73.2 & 74.2 & & 76.1 \\
$B$($E2$;$2^+_2\rightarrow 0^+_1$) & &  7.1 &  7.2 &  7.1 & &  6.8 \\
$B$($E2$;$2^+_2\rightarrow 2^+_1$) & & 12.1 & 16.8 & 16.2 & & 16.6 \\
$B$($E2$;$4^+_1\rightarrow 2^+_1$) & &103.8 &102.6 &102.0 & &101.1 \\
$B$($E2$;$4^+_1\rightarrow 2^+_2$) & &  1.8 &  0.4 &  0.5 & &  0.5 \\ \hline
$Q(2^+_1)$ & & $-$18.7 & $-$18.4 & $-$17.9 & & $-$17.1 \\
$Q(2^+_2)$ & &    18.5 &    18.4 &    18.1 & &    17.3 \\
$Q(4^+_1)$ & & $-$21.1 & $-$21.5 & $-$21.2 & & $-$20.8 \\
\end{tabular}
\end{center}
\end{table}

%\begin{table}
%\caption{Calculated $E2$ transition matrix elements ($e^2$fm$^4$)
% for low-lying states of $^{64}$Ge.
% Effective charges $e_p$=1.33 and $e_n=0.64$ are adopted
% \protect \cite{brown}.
% \label{table2}}
%\begin{center}
%\begin{tabular}{rrrrrr}
%  transition & \multicolumn{5}{c}{$B$($E2$)}  \\ \cline{2-6}
%             & 100 & 200 & 300 & 400 & 500  \\ \cline{1-6}
% $2^+_1\rightarrow 0^+_1$ &  351 & 422 & 471 & 469 & \\
% $2^+_2\rightarrow 0^+_1$ &   11 &   5 &   3 &   3 & \\
% $2^+_2\rightarrow 2^+_1$ &  196 & 359 & 366 & 367 & \\
% $4^+_1\rightarrow 2^+_1$ &  551 & 587 & 753 & 750 & \\
% $4^+_1\rightarrow 2^+_2$ &  111 &  51 &  37 &  34 & \\
%\end{tabular}
%\end{center}
%\end{table}

%\begin{table}
%\caption{Comparison of excitation energies (MeV) 
% between the QMCD and the exact results.
% \label{table3}}
%\begin{center}
%\begin{tabular}{crrrr}
% state & $f_{7/2}$ & $p_{3/2}$ & $p_{1/2}$ & $f_{5/2}$ \\ \hline
%$0^+_1$ & 15.14 & 2.70 & 0.62 & 5.54 \\
%$2^+_1$ & 15.13 & 2.63 & 0.75 & 5.49 \\
%$2^+_2$ & 14.94 & 2.85 & 0.71 & 5.50 \\
%$4^+_1$ & 14.93 & 2.82 & 0.69 & 5.56 \\
%\end{tabular}
%\end{center}
%\end{table}

\begin{figure}
\caption{(a) Energies and (b) expectation values of $J\cdot J$
 of the lowest six states of $^{24}$Mg plotted 
 as a function of the QMCD basis dimension, with 
$N_t=20$ and $\Delta\beta=0.07$ (MeV$^{-1}$).
The exact values are shown by symbols.
Different symbols indicate different angular momenta.
%The abscissa is in the logarithmic scale.
 \label{fig1}}
\end{figure}

\begin{figure}
\caption{Experimental and calculated energy levels of $^{64}$Ge.
The QMCD parameters are $N_t=40$ and $\Delta\beta=0.06$ (MeV$^{-1}$).
The arrows designate $E2$ transitions with $B$($E2$)'s indicated by their widths.
 \label{fig2}}
\end{figure}

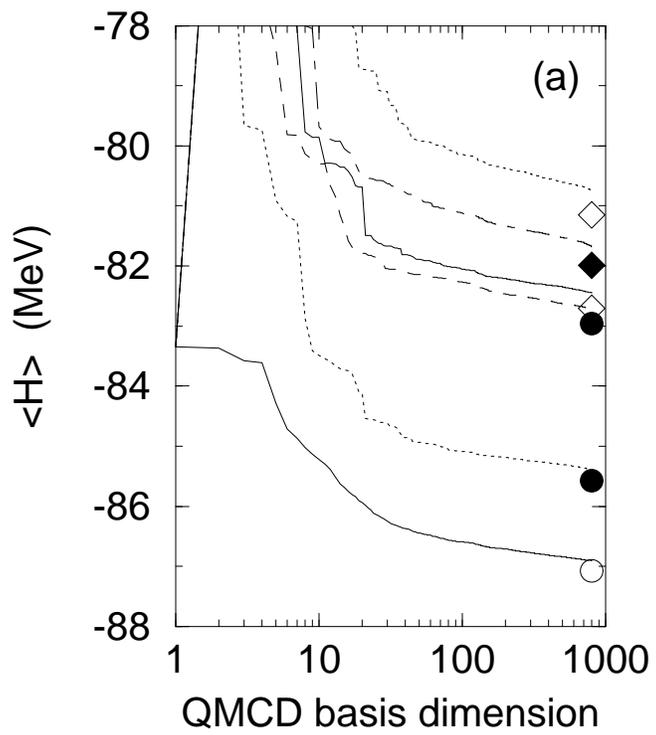
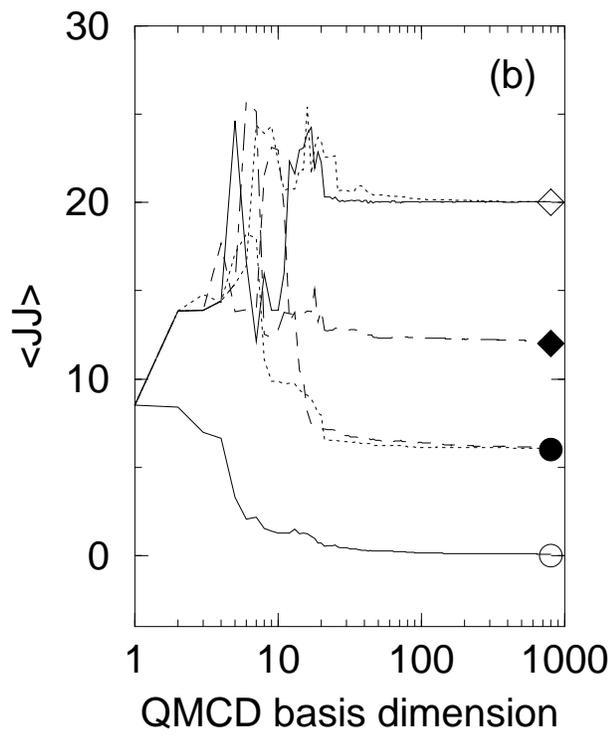
\end{document}